\documentclass[aps,prd,reprint]{revtex4-1}
\usepackage{graphicx}

\begin{document}

\title{Cosmic Microwave Background Delensing Revisited: Residual Biases and a Simple Fix}

\author{Wei-Hsiang Teng}
\affiliation{Department of Physics, National Taiwan University, Taipei 10617, Taiwan} 
\affiliation{Kavli Institute for Particle Astrophysics and Cosmology,
SLAC, 2575 Sand Hill Road, Menlo Park, CA  94025}

\author{Chao-Lin Kuo}
\affiliation{Department of Physics, Stanford University, 382 Via Pueblo Mall, 
Stanford, CA 94305}
\affiliation{Kavli Institute for Particle Astrophysics and Cosmology,
SLAC, 2575 Sand Hill Road, Menlo Park, CA  94025}

\author{Jiun-Huei Proty Wu}
\affiliation{Department of Physics, Institute of Astrophysics, and Center
                      for Theoretical Sciences, National Taiwan University, Taipei 10617, Taiwan}

\begin{abstract}
The delensing procedure is an effective tool for removing lensing-induced $B$-mode polarization in the Cosmic Microwave Background to  allow for deep searches of primordial $B$-modes.  However, the delensing algorithm existing in the literature breaks down if the target  $B$-mode signals overlap significantly with the lensing $B$-mode ($\sim 300<l<2000$) in multipole-$l$ space.  In this paper, we identify  the cause of the breakdown to be correlations between the input $B$-map and the deflection field estimator ($EB$).  The amplitude of  this bias is quantified by numerical simulations and compared to the analytically derived functional form.  We also propose a revised  delensing procedure that circumvents the bias.  While the newly identified bias does not affect the search of degree scale $B$-mode  generated by tensor perturbations, the modified delensing algorithm makes it possible to perform deep searches  of high-$l$ $B$-modes such as those generated by patchy reionization, cosmic strings, or rotation of polarization angles.  Finally, we  estimate how well future polarization experiments can do in detecting tensor- and cosmic string- generated $B$-mode after delensing and comment on different survey strategies.  
\end{abstract}
\maketitle

\section{Introduction}
Linear, scalar perturbations do not generate $B$-mode polarization in the cosmic microwave background (CMB) \cite{kami97} \cite {selj97}. This exact symmetry makes $B$-mode polarization a power tool to look for ingredients beyond standard cosmology, such as  tensor perturbations (gravitational waves), cosmic strings, patchy reionization, and polarization rotation \cite{selj06} \cite{pogo08}  \cite{arva10} \cite{kami09} \cite{dvor09}. Gravitational lensing by large scale structures converts $E$-mode to $B$-mode, acting as a  foreground contamination.  Because the frequency dependence of lensing $B$-mode is that of a 2.73K blackbody, it cannot be removed with  multi-frequency observations and will eventually limit future $B$-mode observations \cite{knox02} \cite{kes02}.

Fortunately, because of the unique non-Gaussian signatures in the lensing $B$-mode, it can be separated from the primordial Gaussian  $B$-mode \cite{zalda98} \cite{zalda00}.   This ``delensing" procedure reconstructs the intervening lens by looking at the higher order statistics in the temperature  and polarization maps \cite{hu01} \cite{hu02} \cite{hira03}, calculates the expected lensing $B$-mode from the observed $E$-mode and the reconstructed mass distribution, and  subtracts it from the observed $B$-mode \cite{smith08}.  It has been shown that this procedure is effective in removing the lensing $B$-mode \cite{selj04}.   External data sets can be used to gain information on the matter distribution along the line of sight, but if the instrument noise is  low ($<\sim 3 \mu $K\,arcmin) \cite{smith10}, internal delensing from the CMB becomes much more efficient than delensing with external  data.

In this paper, we point out that because of higher order effects that have been ignored in previous treatments, the delensing algorithm  as described in the literature fails if the target $B$-mode signals reside in the same $l$ range as the lensing $B$-mode ($\sim  300<l<2000$).  This does not create a major problem for the search of degree scale $B$-mode generated by tensor perturbations, because  the observed $B$-mode map can be high-pass filtered (above $l\sim 150$) before the lensing reconstruction \cite{selj04}.  However, the  current delensing algorithm is not applicable to the search of high-$l$ $B$-modes such as those generated by patchy reionization,  cosmic strings, or rotation of polarization angles. In this paper, we identify the cause of the breakdown to be a sizable correlation  between the input $B$-map and the deflection field $EB$ estimator.  This is important because at the sensitivity level achievable by  next generation ground based experiments, the most sensitive estimator for the deflection field is the $EB$ estimator \cite{hu02}.  We  explicitly derive the functional forms for a few of the bias terms and verify them with numerical simulations.  The phenomenology caused by  the bias is also explored and discussed.  

In section \ref{sec:debias}, we propose a revised delensing procedure that takes cares of all the high order terms that we have not analytically  calculated, and show in simulations that it avoids the bias.  This method is straightforward to implement, and manifestly bias-free.  Furthermore, the new algorithm can be applied to quadratic estimators as well as likelihood-based lensing estimators  (iterative estimators), allowing the lowest possible residuals to be achieved.  In section \ref{sec:forecast}, we apply the results obtained in  section \ref{sec:debias} to forecast the sensitivity of future experiments to tensor-induced and cosmic string-induced $B$-polarization  with iterative delensing.   

Throughout this paper, We use a flat $\Lambda$CDM cosmology with parameters $\Omega_b h^2=0.0226$, $\Omega_c h^2=0.11$, $h=0.7$,  $n_s=0.96$, $\Delta^2_{\mathcal{R}}(k_0)=2.46\times 10^{-9}$ at $k_0=0.002$ Mpc$^{-1}$, and $\tau=0.09$. The tensor-to-scalar ratio is defined as initial tensor/scalar power spectrum amplitude: $r=\Delta^2_{h}(k_0)/\Delta^2_{\mathcal{R}}(k_0)$.  The tensor and lensing $B$-mode  power spectra are produced by CAMB \cite{lewis00}, and the string $B$-mode power spectrum is using CMBACT \cite{pogo99}.  All  simulations were carried out under the flat-sky approximation.  In this paper, we focus on the treatment of delensing bias and defer  the complete treatment of delensing with $EB$ leakage from map boundaries to a future paper.  

\begin{figure}
\centering
\includegraphics[width=8.6cm]{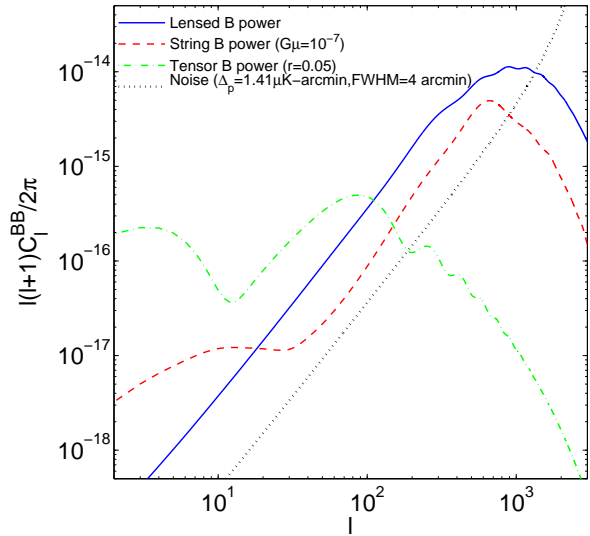}
\caption{This figure shows the shapes of $B$-mode power spectra generated by tensor perturbation (tensor-to-scalar ratio $r=0.05$),  cosmic strings (strings tension $G\mu=10^{-7}$) and the gravitational lensing. While the tensor-induced $B$-mode and lensing $B$-mode  are well separated in $l$ space, the lensing $B$-mode overlaps with $B$-mode from strings and cannot be delensed using high-pass  filters.  The same is true for rotation generated $B$-mode \cite{arva10}, which would have an $E$-mode like spectrum that extends to high-$l$  (not shown). }
\label{fig:plot0}
\end{figure}

\begin{figure}
\centering
\includegraphics[width=8.6cm]{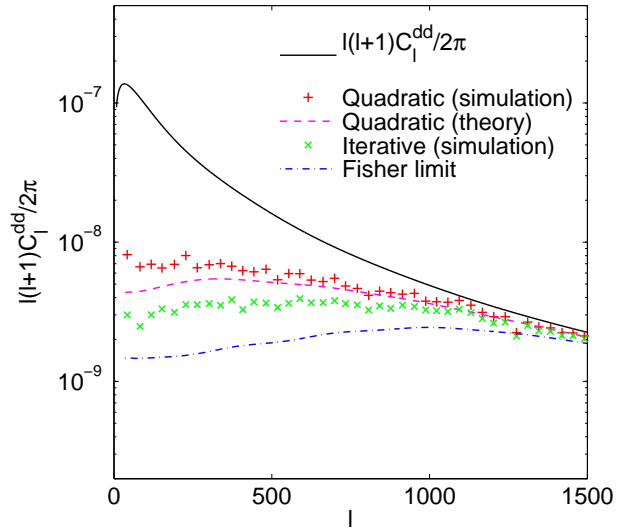}
\caption{The comparison of simulated reconstruction error (or bias on power spectrum of deflection field) with theory using quadratic  and iterative estimators ($EE$ and $EB$).  The theoretical estimate of error and Fisher limit are from Eq. (66) and (32) in \cite {hira03}.  The instrument noise level is set to be $\Delta_p=1.41\mu$K\,arcmin in these simulations, with FWHM$=4'$.  Here we use 64  realizations of CMB fields but only one realization of lensing potential.  We choose the convergence factor to be $0.12$ and perform 64  iterations, as suggested in \cite{hira03}.  }
\label{fig:plot1}
\end{figure}

\section{lens reconstruction}

In this section we provide a brief overview of gravitational lensing and deflection field reconstruction. For more details please refer to \cite{hu02} \cite{hira03} \cite{smith10}. 

The large scale structures deflect the CMB photons randomly as they propagate through the universe. Since the deflection angles are small ($\sim$ 2 arcmin), we consider only the first order density perturbations and apply the Born approximation.  We can then model the lensed fields as: 

\begin{eqnarray}
\label{eqn:lens}
&&X(\hat{{\bf n}})=\tilde{X}(\hat{{\bf n}}+d(\hat{{\bf n}}))
\end{eqnarray}
$X(\hat{{\bf n}})$ can be CMB temperature $T$ or Stokes parameters $Q$ and $U$. The tilde indicates the unlensed fields and $d(\hat {{\bf n}})$ is the deflection angle which is the gradient of integrated lensing potential $\phi(\hat{{\bf n}})$ (for more detail see  \cite{lewis06}).

The gravitational lensing breaks the statistical isotropy in the unlensed CMB.  In fact, this is what makes the reconstruction of the  deflection field possible.  The quantity $\nabla\cdot(\bar{X} \nabla \hat{X})$ gives us the best quadratic estimator to detect the anisotropy caused by lensing  \cite{hu02}, where $\bar{X}$ is the filtered field $(C^{XX}+C^{NN})^{-1}X$.  The quantity $C^{NN}$ is the  noise covariance matrix, and $C^{XX}$ is the covariance matrix of the lensed field averaged over lensing potential realizations so that  it is diagonal in harmonic space.  $\hat{X}$ is another filtered field $C^{\tilde{X}\tilde{X}}(C^{XX}+C^{NN})^{-1}X$, where $C^{\tilde {X}\tilde{X}}$ is the covariance matrix of unlensed field.  

A more complicated likelihood-based method that goes beyond the quadratic order has also been introduced to improve the lens  reconstruction \cite{hira03}.  This is especially relevant for future CMB experiments that have low instrument noise (a few $\mu$K\,arcmin).  The main  distinction between this approach and the quadratic estimator is that the likelihood method replaces the weighting $(C^{XX}+C^{NN})^{- 1}$ by $(C^{\tilde{X}\tilde{X}}+C^{NN})^{-1}$ and introduces mean field subtraction.  As pointed out in \cite{han09}, this also  provides a way to deal with non-uniform sky coverage.  The likelihood-based method can also be understood as an iterative process that  recalculates the quadratic estimator repeatedly.  The lensing-induced $B$-modes can be removed iteratively until the series converges  to a residual $B$-mode map that depends on instrument noise level.  

We developed a simulation pipeline based on the procedures outlined above and tested it with a variety of input parameters.  We  followed the algorithms in \cite{hu02} and \cite{hira03} to simulate $EE$ and $EB$ quadratic and likelihood-based (iterative)  estimators and obtained results consistent with those in their papers (see Fig. \ref{fig:plot1}).  Throughout this paper, our simulations  focus on noise level: $\Delta_p=1.41\mu$K\,arcmin, FWHM$=4'$ and sky coverage: $10^{\circ}\times10^{\circ}$, with grid spacing of 1 arcmin, unless  otherwise stated.  This noise level is achievable in future ground-based experiments such as the POLAR Array.  It is worth mentioning that the analytic estimate for the quadratic estimator error  \cite{hu02} is lower than the simulated results.  This is the result of approximating $E$ and $B$ modes as Gaussian, a known effect reported in \cite{coo03} \cite{kes03} \cite{han10}.  The analytic estimate for the reconstruction uncertainty in iterative estimator is more difficult.  Here we plot the Fisher limit to provide the lower bound of the lens reconstruction error (see \cite{hira03}  for details).

\section{Delensing and the Cause of the Delensing Bias}
\label{sec:del}

After reconstructing the lensing potential, or equivalently the deflection field, we can use this information to either numerically  reverse lensing to obtain unlensed $TQU$ maps, or to calculate the expected $B$-polarization induced by lensing and subtract it  from the observed $B$-map.  We tried both methods in simulations and did not find significant difference between the two at low $l$  region, before the effects of pixelization kicks in.  We choose the latter method (subtraction-based) for most of the simulations since  it has a nice analytic representation in Fourier space.  In practice, we Wiener-filter the observed $E$-mode (lensed and with  instrumental noise) and approximate it as the unlensed $E$-mode.  This approximation is justified because weak lensing only alters $E$-mode slightly.  We then apply the deflection field estimator to ``lens'' this $E$-map to calculate the lensing induced $B$-mode.  To  first order in lensing potential $\phi$, the expected $B$-mode from lensing is given by 

\begin{eqnarray}
\label{eqn:elens}
\hat{B}^{\rm lens}(l_1)=\int\frac{d^2l_2}{(2\pi)^2}f(l_1,l_2)\frac{C^{EE}_{l_2}E^{\rm obs*}(l_2)}{C^{EE}_{l_2}+N^{EE}_{l_2}}\frac{C^ {\phi\phi}_l\hat{\phi}(l)}{C^{\phi\phi}_l+N^{\phi\phi}_l}, \nonumber \\
\end{eqnarray}

$N^{\phi\phi}_l$ is lens reconstruction error. For $EB$ estimator and approximating $E$ and $B$ modes as Gaussian we can express it as:

\begin{equation}
\label{eqn:recnoise}
N^{\phi\phi}_l=\left[\int\frac{d^2l_1}{(2\pi)^2}\frac{f(l_2,l_1)^2 (C^{EE}_{l_1})^2}{(C^{E_{\rm lens}}_{l_1}+N^{EE}_{l_1})(C^{B_{\rm  lens}}_{l_2}+N^{BB}_{l_2})}\right]^{-1},
\end{equation}

where $f(l_1,l_2)=sin2\varphi_{l_1l_2}(l_2\cdot l)$ ($l=l_1+l_2$), $\hat{\phi}$ is the estimated lensing field, $C^{E_{\rm lens}}_l$ and $C^{B_{\rm lens}}_l$ are the lensed $EE$ and $BB$ power spectra.  $\varphi_{l_1l_2}=\varphi_{l_1}-\varphi_{l_2}$ ($\varphi_l$ is angle between $l$ and x-axis). This estimator is exactly Eq.(5) in  \cite{smith10} under flat sky approximation.  The estimator for delensed $B$-mode is simply the difference between the observed and  estimated $B$ mode
\begin{equation}
\label{eqn:edel}
\hat{B}^{\rm del}(l)=B^{\rm obs}(l)-\hat{B}^{\rm lens}(l).
\end{equation}

The observed $B$-mode $B^{\rm obs}(l)$ includes $B$-mode from lensing $B^{\rm lens}(l)$, noise $B^{\rm noise}(l)$, and primordial $B$-mode $B^{\rm pri}(l)$ if it exists. The primordial power spectrum based on squaring this estimator ($\hat{C}^{B_{\rm del}} _l=\langle\hat{B}^{\rm del*}(l)\hat{B}^{\rm del}(l)\rangle$) will be biased.  The sources of biases are imperfect lens reconstruction,  instrument noise in the observed $E$, $B$ maps, and their complicated cross power spectra. Out of the 10 quadratic terms in  $\hat{C}^{B_{\rm del}}_l$, we know $B^{\rm lens}(l)$, $B^{\rm noise}(l)$ and $B^{\rm pri}(l)$ do not correlate with one another. As a result, there are only 7 non-trivial terms.  Two of them are auto-spectra, primordial 
$B$-mode power $C^{\rm B_{pri}}_l$ and noise power $C^{\rm B_{noise}}_l$. We denote the total contribution from the rest of the 5 terms as $C^{B_{\rm res}}_l$.  Explicitly,

$$
C^{B_{\rm res}}_l=C^{B_{\rm lens}}_l+\langle \hat{B}^{\rm lens*}(l)\hat{B}^{\rm lens}(l)\rangle-2\langle B^{\rm lens*}(l)\hat{B}^{\rm lens}(l)\rangle
$$
\begin{equation}
-2\langle B^{\rm pri*}(l)\hat{B}^{\rm lens}(l)\rangle-2\langle B^{\rm noise*}(l)\hat{B}^{\rm lens}(l)\rangle.\label{eqn:crosspower}
\end{equation}

\begin{figure*}
\centering
\includegraphics[width=8.6cm]{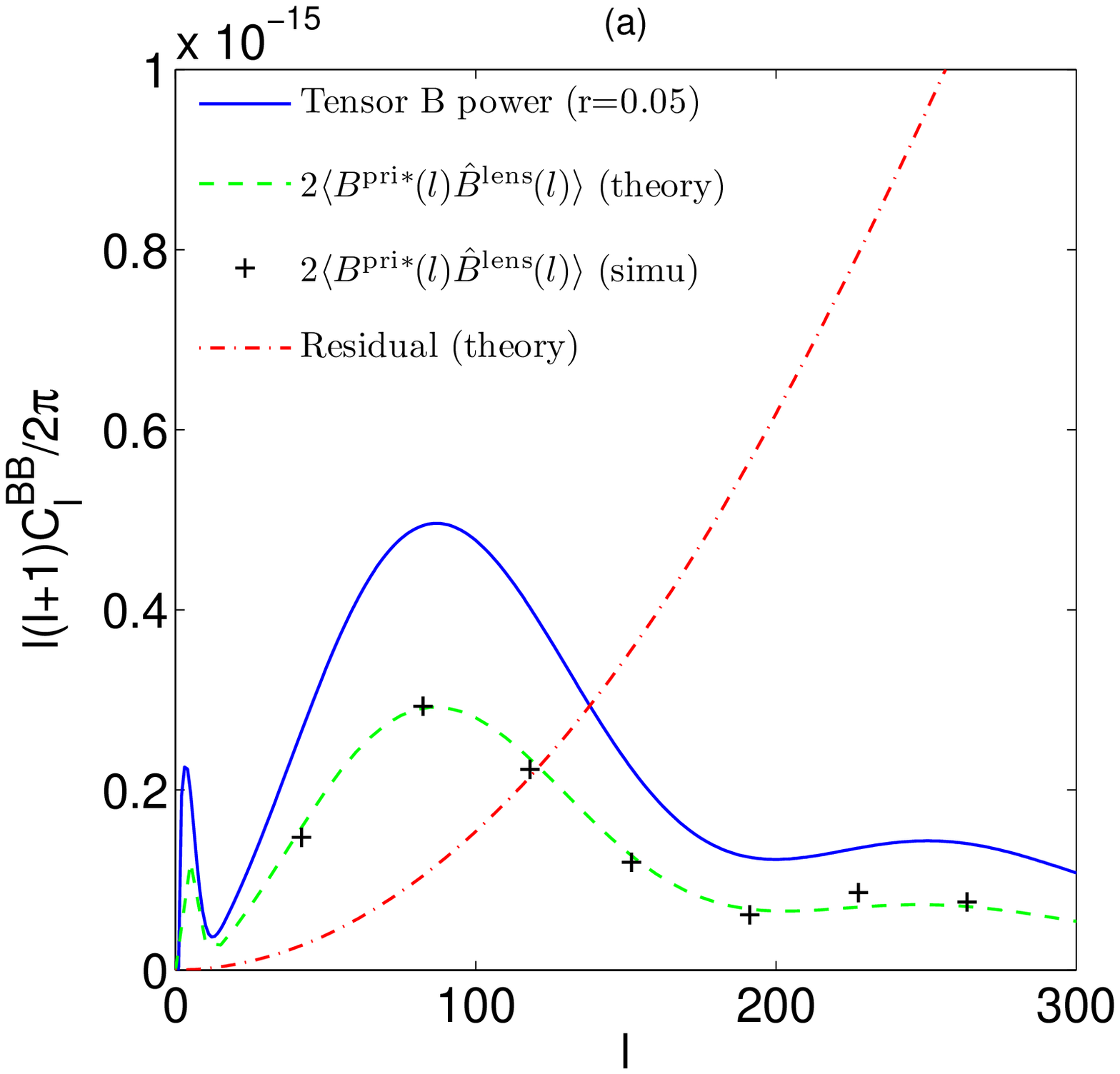}
\includegraphics[width=8.6cm]{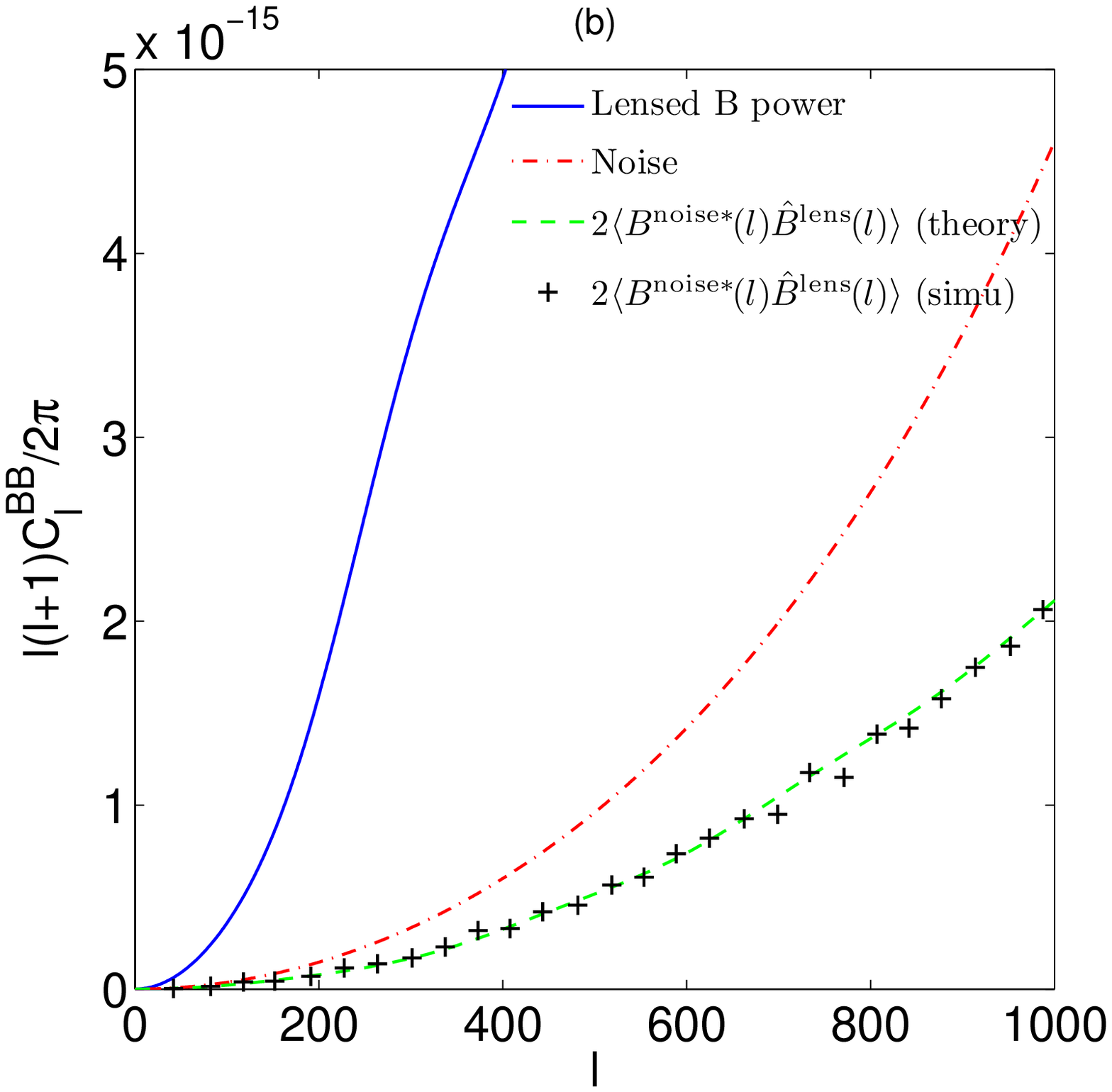}
\caption{These plots demonstrate the delensing biases caused by high order correlations. {\it (a):} the blue line is tensor $B$ -mode power ($r=0.05$) and the red dot-dashed line is delensing residuals predicted by Eq. (\ref{eqn:res}), first given in \cite {smith10}. The green dashed line is theoretical prediction of $2\langle B^{\rm pri*}(l)\hat{B}^{\rm lens}(l)\rangle$ given by Eq. (\ref{eqn:bias}). The black crosses are simulation result using an $EB$ quadratic estimator (400 realizations). This shows that the bias is far from negligible compared to the residual. {\it  (b):} The blue line is lensing $B$ power and the red dot-dashed line is noise power ($\Delta_p=1.41\mu$K\,arcmin, FWHM=4 arcmin). the green dashed line is theoretical prediction of $2\langle B^{\rm noise*}(l)\hat{B}^{\rm lens}(l)\rangle$ given by Eq. (\ref{eqn:biasn}). The black crosses are simulation result using an $EB$ quadratic estimator (400 realizations). Again, the bias is almost 60$\%$ of the total residual. }
\label{fig:plot2}
\end{figure*}

All these terms will contribute to the lensing residuals, including the negative contributions from the cross correlations.  Knowing the delensing biases for a given set of experimental parameters is important for planning of future CMB experiments.   Furthermore, the properties of the residuals, including their correlations, need to be understood in actual data analysis so that  significance of the target signals can be evaluated.  An approximated residual for an $EB$ quadratic estimator has been given analytically by \cite{smith10}:

\begin{eqnarray}
\label{eqn:res}
C^{B_{\rm res}}_l= \;\;\;\;\;\;\;\;\;\;\;\;\;\;\;\;\;\;\;\;\;\;\;\;\;\;\;\;\;\;\;\;\;\;\;\;\;\;\;\;\;\;\;\;\;\;\;\;\;\;\;\;\;\;\;\;\;\;\;\;\;\;\;\;\;\;\;\;\;\;\;\;\;\;\;\;  \nonumber  \\
\int\frac{d^2l_1}{(2\pi)^2}f(l,l_1)^2\left[C^{\phi\phi}_{l_2}C^{EE}_{l_1}-\frac{(C^{\phi\phi}_{l_2})^2}{C^{\phi\phi}_{l_2}+N^ {\phi\phi}_{l_2}}\frac{(C^{EE}_{l_1})^2}{C^{EE}_{l_1}+N^{EE}_{l_1}}\right], \nonumber \\
\end{eqnarray}
where $l_2=l+l_1$. This estimated residual apparently came from the first three terms in Eq. (\ref{eqn:crosspower}), representing imperfect lens reconstruction ($N^{\phi\phi}_{l}$) and instrument noise ($N^ {EE}_{l}$). As these two terms vanish, the residual vanishes as well.  In deriving this residual \cite{smith10} ignores $2\langle B^{\rm pri*}(l)\hat{B}^{\rm lens}(l)\rangle$ and $2\langle B^{\rm noise*}(l)\hat{B}^{\rm lens}(l)\rangle$, even though they admited it has not been justified. 
They also approximated the calculation of $\langle \hat{B}^{\rm lens*}(l)\hat{B}^{\rm lens}(l)\rangle$ and $\langle B^{\rm lens*}(l)\hat{B}^{\rm lens}(l)\rangle$ by neglecting $\langle E^{\rm obs}E^{\rm obs}\phi\phi\rangle_{\rm c}$ connected four-point function.  

The residual for iterative estimator can be evaluated in the same way using Eq. (\ref{eqn:recnoise}) and Eq. (\ref{eqn:res}). The only difference is iterative estimator improves the ability of lens reconstruction by depressing $C^{B_{\rm lens}}_l$ to $C^{B_{\rm res}}_l$. Therefore, $C^{B_{\rm res}}_l$ will appear simultaneously in the left-hand and right-hand sides of Eq. (\ref{eqn:res}). We can use an iterative algorithm to solve it.

\section{the magnitude of delensing bias}

In this section, we show the neglected terms ( $\langle B^{\rm pri*}(l)\hat{B}^{\rm lens}(l)\rangle$, $\langle B^{\rm noise*}(l)\hat{B}^{\rm lens}(l)\rangle$ and $\langle E^{\rm obs}E^{\rm obs}\hat{\phi}\hat{\phi}\rangle_{\rm c}$ ) contribute significantly to $C^{B_{\rm res}}_l$ under parameters that are typical of future experiments. In the simplest case, we obtain $\hat{\phi}$ from $EB$ quadratic estimator and the estimated lensed $B$-mode is quadratic estimator of $E^{\rm obs}$ and $\hat{\phi}$, $\langle B^{\rm pri*}(l)\hat{B}^{\rm lens}(l)\rangle$ will be four-point function $\langle B^{\rm pri}E^{\rm obs}B^{\rm obs}E^{\rm obs}\rangle$. By applying Wick's theorem and ignoring $EB$ cross correlation, we obtain

\begin{eqnarray}
\label{eqn:bias}
\langle B^{\rm pri*}(l)\hat{B}^{\rm lens}(l) \rangle=\;\;\;\;\;\;\;\;\;\;\;\;\;\;\;\;\;\;\;\;\;\;\;\;\;\;\;\;\;\;\;\;\;\;\;\;\;\;\;\;\;\;\;\;\;\;\;\;\;\;\;\;  \nonumber \\
\frac{C^{B_{\rm pri}}_l}{C^{B_{\rm lens}}_l+N^{BB}_l}\int\frac{d^2l_1}{(2\pi)^2}\frac{f(l,l_1)^2(C^{EE}_{l_1})^2}{C^{EE}_{l_1}+N^{EE}_ {l_1}}\frac{C^{\phi\phi}_{l_2}N^{\phi\phi}_{l_2}}{(C^{\phi\phi}_{l_2}+N^{\phi\phi}_{l_2})} ,\nonumber \\
\end{eqnarray}
where $l_2=l+l_1$. Here we ignore the non-scalar $E$-mode because it will be much smaller than scalar $E$-mode. In practice, the bias term contributes negatively to the delensed power Eq. (\ref{eqn:crosspower}) with an $l$ dependence similar to the primordial $B$-mode.  

Since $B^{\rm noise}(l)$ acts like $B^{\rm pri}(l)$ in the cross correlation, the cross correlation $\langle B^{\rm noise*}(l)\hat{B}^{\rm lens}(l)\rangle$ is in the form similar to Eq. (\ref{eqn:bias}) except that $C^{B_{\rm pri}}_l$ is replaced by $N^{BB}_l$,

\begin{eqnarray}
\label{eqn:biasn}
\langle B^{\rm noise*}(l)\hat{B}^{\rm lens}(l) \rangle=\;\;\;\;\;\;\;\;\;\;\;\;\;\;\;\;\;\;\;\;\;\;\;\;\;\;\;\;\;\;\;\;\;\;\;\;\;\;\;\;\;\;\;\;\;\;\;\;\;\;\;\;  \nonumber \\
\frac{N^{BB}_l}{C^{B_{\rm lens}}_l+N^{BB}_l}\int\frac{d^2l_1}{(2\pi)^2}\frac{f(l,l_1)^2(C^{EE}_{l_1})^2}{C^{EE}_{l_1}+N^{EE}_ {l_1}}\frac{C^{\phi\phi}_{l_2}N^{\phi\phi}_{l_2}}{(C^{\phi\phi}_{l_2}+N^{\phi\phi}_{l_2})} .\nonumber \\
\end{eqnarray}

Our simulation results (Fig.\ref{fig:plot2}) show good consistency with the analytic calculation. Up to $\sim 60\%$ of primordial $B$-mode power spectrum will be supressed (at $l\sim 60$) by the cross correlation term $-2\langle B^{\rm pri*}(l)\hat{B}^{\rm lens}(l)\rangle$ (the location of recombination peak of tensor $B$-mode). Even the instrumental noise is artificially ``delensed" by a similar fraction because of the contribution from $-2\langle B^{\rm noise*}(l)\hat{B}^{\rm lens}(l)\rangle$ in Eq. (\ref{eqn:crosspower}). As a result, delensing simply fails to improve signal-to-noise ratio of the primordial $B$-mode.

\begin{figure}
\centering
\includegraphics[width=8.6cm]{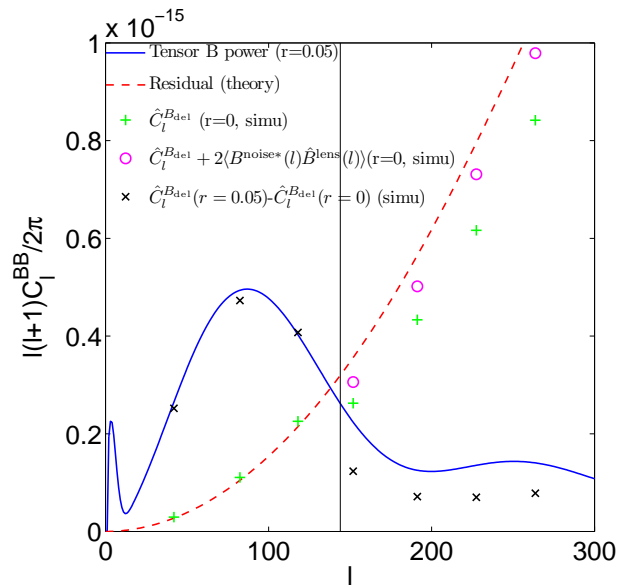}
\caption{The simulation result with cutoff at $l=144$ (black vertical solid line) for $r=0.05$ and $r=0$.  At low $l$, the residual power recovers the theoretical prediction (sum of instrumental noise and reconstruction noise, bias-free), while at high $l$ there are  contributions from terms such as $\langle E^{\rm obs}E^{\rm obs}\hat{\phi}\hat{\phi}\rangle_{\rm c}$ and $\langle B^{\rm noise*} (l)\hat{B}^{\rm lens}(l)\rangle$ for $r=0$, or $\langle B^{\rm pri*} (l)\hat{B}^{\rm lens}(l)\rangle$ for $r=0.05$.}
\label{fig:plot2_1}
\end{figure}

The calculation of the additional bias term $\langle E^{\rm obs}E^{\rm obs}\phi\phi\rangle_{\rm c}$ is daunting, if $\phi$ is obtained by an $EB$ estimator.  It will be an 8-point function $\langle E^{\rm obs}E^{\rm  obs}E^{\rm obs}E^{\rm obs}E^{\rm pri}E^{\rm pri}\phi\phi \rangle$. Therefore, we choose to do simulation to see if $2\langle B^{\rm noise*}(l)\hat{B}^{\rm lens}(l)\rangle$ can account for all the biases in estimating $\hat{C}^{B_{\rm del}}_l$, in the case of zero tensor $B$-mode. In Fig.\ref{fig:plot2_1}, We follow the delensing procedure described in section \ref{sec:del} to get $\hat{C}^{B_{\rm del}}_l$ for both cases with $r=0.05$ and $r=0$. We have removed $E$ and $B$ modes at $l<144$ in lens reconstruction. We can see that even though $\hat{C}^{B_{\rm del}}_l$ has been corrected by the simulated $2\langle B^{\rm noise*}(l)\hat{B}^{\rm lens}(l)\rangle$, the resulting $\hat{C}^{B_{\rm del}}_l$ is still lower than $C^{B_{\rm res}}_l$ estimated in Eq.  (\ref{eqn:res}).  This indicates that the contribution from $\langle E^{\rm obs}E^{\rm obs}\phi\phi\rangle_{\rm c}$ is not neglegible. 

All these biases and correlations can be avoided for searches of tensor-induced $B$-mode, because the modes located at low $l$ can  simply be removed prior to lensing reconstruction.  As a result, $\hat{\phi}$ contains no $E^{\rm obs}$ and $B^{\rm obs}$ at low $l$  region where tensor $B$ resides and where $\hat{B}^{\rm lens}_l$ matters.  This is what has been done in \cite{selj04}.  Under this  condition Eq. (\ref{eqn:res}) is a good estimate until the noise level is so low that the effects of neglecting the non-Gaussianity in  $N^{\phi\phi}_l$ become present.  
 
As we pointed out earlier, this method obviously breaks down when one applies delensing to the search of $B$-mode at high $l$ (such as  cosmic strings or rotation induced $B$-mode).  The biases are far from negligible even for a modest noise level.  One could try to  calculate all the bias terms and correct for them.  However, it is difficult to analytically track all the bias terms in the quadratic  estimator.  The situation is even more complicated for the iterative estimator because of all the N-point functions introduced by  iterating.  Even if an expression is found for all the correction terms, their numerical evaluations can also be non-trivial.  In the  next section, we propose a revised delensing process that works at high $l$. 

\section{A Bias-free Delensing Procedure} 
\label{sec:debias}

\begin{figure}
\centering
\includegraphics[width=8.6cm]{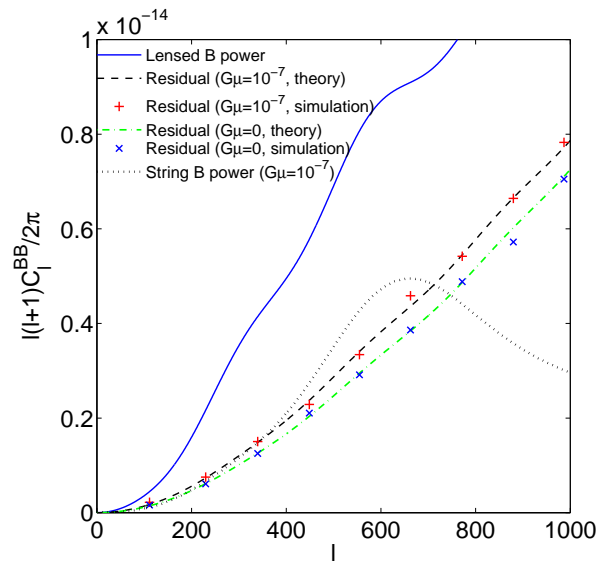}
\caption{This figure is the main result of this paper.  It shows the comparison of theoretical predictions and simulated delensing  residuals using a joint $EE$ and $EB$ iterative estimator.  With the ``moving notch filtering'' described in the text, the simulated delensing  residuals are in excellent agreement with the theoretical predictions calculated from Eq. (\ref{eqn:res}) without the biases described in section IV.  For demonstrative purposes, we assume a string induced primordial $B$-mode $G\mu=10^{-7}$).  When the primordial $B$-mode is non-zero, we replace $C^{B_{\rm  lens}}_l$ in Eq. (\ref{eqn:res}) with $C^{B_{\rm lens}}_l+C^{B_{\rm pri}}_l$ to predict the theoretical residuals.  The annuli are chosen  to be $l=0-180$ and $l=180-1044$ with spacing equals $108$.  Each simulated point contains 25 CMB realizations.}
\label{fig:plot3}
\end{figure}

We choose not to remove the delensing biases by explicitly calculating all the terms.  Instead, we adopt a similar method used in  \cite{selj04}, in which the measured $B$-map is high-pass filtered before the deflection field reconstruction.  In \cite{selj04} the  $l$-cut was chosen to be at $l\sim 150$, leaving most of the lensing information intact.  The cutoff ensures the instrumental noise at  low $l$ and the degree-scale primordial $B$-mode are not artificially ``delensed" due to the biases discussed in the previous section  (Fig \ref{fig:plot2}).  Realizing that, we can revise this method for high $l$ region.  We proposed a modified estimator $\bar{B}^{\rm  del}(l)$ instead of Eq. (\ref{eqn:edel}):

\begin{eqnarray}
\label{eqn:moddel}
\bar{B}^{\rm lens}_N(l_1)=\int\frac{d^2l_2}{(2\pi)^2}f(l_1,l_2)\frac{C^{EE}_{l_2}E^{\rm obs*}(l_2)}{C^{EE}_{l_2}+N^{EE}_{l_2}}\frac{C^ {\phi\phi}_l\bar{\phi}_N(l)}{C^{\phi\phi}_l+N^{\phi\phi}_l} \nonumber \\
\bar{B}^{\rm del}(l)=B^{\rm obs}(l)-\sum_{N}\bar{B}^{\rm lens}_N(l)  \;\;\;\;\;\;\;\;\;\;\;\;\;\;\;\;\;\;\;\;\;\;\;\;\;\;\;\;\;\;\nonumber \\
\end{eqnarray} 
First, the Fourier space is divided into several non-overlapping annuli.  The annuli are labeled by index $N=1,2...$.  The area of the annulus determines how many modes we lose in lens reconstruction.  Instead of high-pass filtering, we implement a ``notch filter" to the Fourier $B$-map for each region $N$.  In other words, the modified estimator of lensing potential in region $N$, $\bar{\phi}_N (l)$, would be the ordinary quadratic or iterative estimators except that the modes located {\it inside} the annulus $N$ are set to  zero.  As usual, the estimator for lensed $B$-mode, $\bar{B}^{\rm lens}_N(l)$, will be calculated using $\bar{\phi}_N(l)$,  except that the result of $\bar{B}^{\rm lens}_N(l)$ {\it outside} the annulus $N$ will be ignored because they have been contaminated  by delensing bias.  Repeat the process for a different annulus, eventually we obtain uncontaminated $\bar{B}^{\rm lens}_N(l)$  corresponding to each annulus.  The last step is straightforward.  We compute the delensed $B$-mode $\bar{B}^{\rm del}(l)$ by  subtracting the combination of these uncontaminated estimated lensed $B$-modes from the observed $B$-mode.  The estimated power  spectrum from $\bar{B}^{\rm del}(l)$ is now free of the delensing biases. In Fig.\ref{fig:plot3}, we show that this procedure avoids  the delensing biases and the delensing residuals now agree with theoretical expectations very well.  

Ignoring the $B$-map in the annulus where delensing takes place will obviously degrade the signal-to-noise.  The reduction in signal-to-noise can be minimized by dividing the $B$-map into smaller annuli, if one is willing to pay the price of proportionally increased computing time.  This is still considerably simpler compared to a more analytic treatment of $\langle\hat{B}^{\rm lens*}(l)\hat{B}^ {\rm lens}(l)\rangle$, which will have to be computed numerically anyway.  Another significant advantage is, it is straightforward to  implement this ``moving notch filter" approach to iterative estimators to have the lowest possible delensing residuals (Fig. \ref {fig:plot3}).  

It is known that the cut sky effects make the neighboring modes in Fourier space correlated.  In this case the removal of $E$, $B$  modes in certain annulus described above would not guarantee perfectly avoidance of delensing biases.  We might improve the behavior by  extending these annuli to overlap (each side to $\Delta l \sim 360/\theta$) but preserve the modes $\bar{B}^{\rm lens}_N(l)$ in the  original annulus.  We see this as yet another issue related to polarization delensing with map boundaries (in addition to other  unresolved issues) and defer a full treatment to future work.    

With the ``moving notch filter'' approach, the residual is now free of the correlation terms that are plaguing the delensing process. However, to estimate the primordial $B$ power, the residual power still needs to be subtracted from the delensed $B$-power estimator. If primordial $B$-mode exists, it will contribute to lens reconstruction error $N^{\phi\phi}_l$. In other words, the residual power is understandably a function of the primordial $B$ power. This is clearly seen in Fig. \ref{fig:plot3}.  This issue can be easily resolved with a standard iterative process.  We start by assuming there is no primordial $B$ and subtract the appropriate residual from the delensed $B$ power estimator. If the estimated primordial $B$ agrees with zero all is good. If not, we iterate the formula below until it converges:
\begin{equation}
\label{eqn:prim}
\hat{C}^{B_{\rm pri}}_l=\langle B'^{\rm del*}(l)B'^{\rm del}(l)\rangle - C^{B_{\rm res}}_l(\hat{C}^{B_{\rm pri}}_l),
\end{equation}
where $B'^{\rm del}(l)$ is the same as Eq. (\ref{eqn:moddel}), except that we alter the weighting in lens reconstruction by including a primordial $B$-mode $\hat{C}^{B_{\rm pri}}_l$. The second term $C^{B_{\rm res}}_l(\hat{C}^{B_{\rm pri}}_l)$ is Eq. (\ref{eqn:res}) except we replace $C^{B_{\rm lens}}_l$ by $C^{B_{\rm lens}}_l+C^{B_{\rm pri}}_l$ in Eq. (\ref{eqn:recnoise}) when evaluating $N^{\phi\phi}_l$.  $C^ {E_{\rm lens}}_l$ remains the same because primordial $E$-mode is much smaller than lensed $E$-mode.  In actual data analysis we need to have a precise prediction for ``residual'' and understand the convergence properties. In Fig. \ref{fig:plot3} we demonstrated percent level agreements between theory and simulations.  
     
\section{Application of Delensing for tensor and cosmic string generated B-mode}
\label{sec:forecast}

\begin{figure}
\centering
\includegraphics[width=8.6cm]{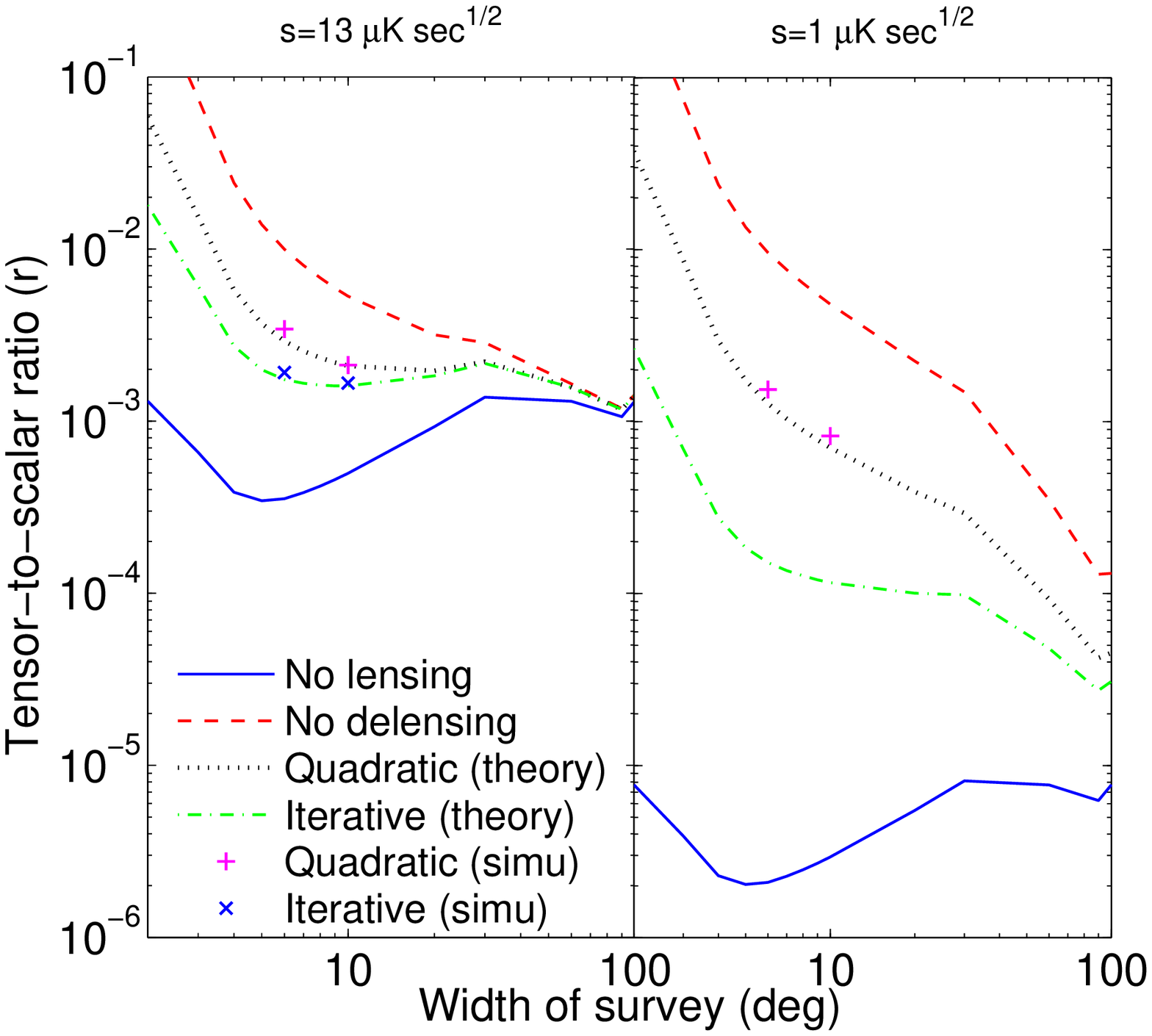}
\includegraphics[width=8.6cm]{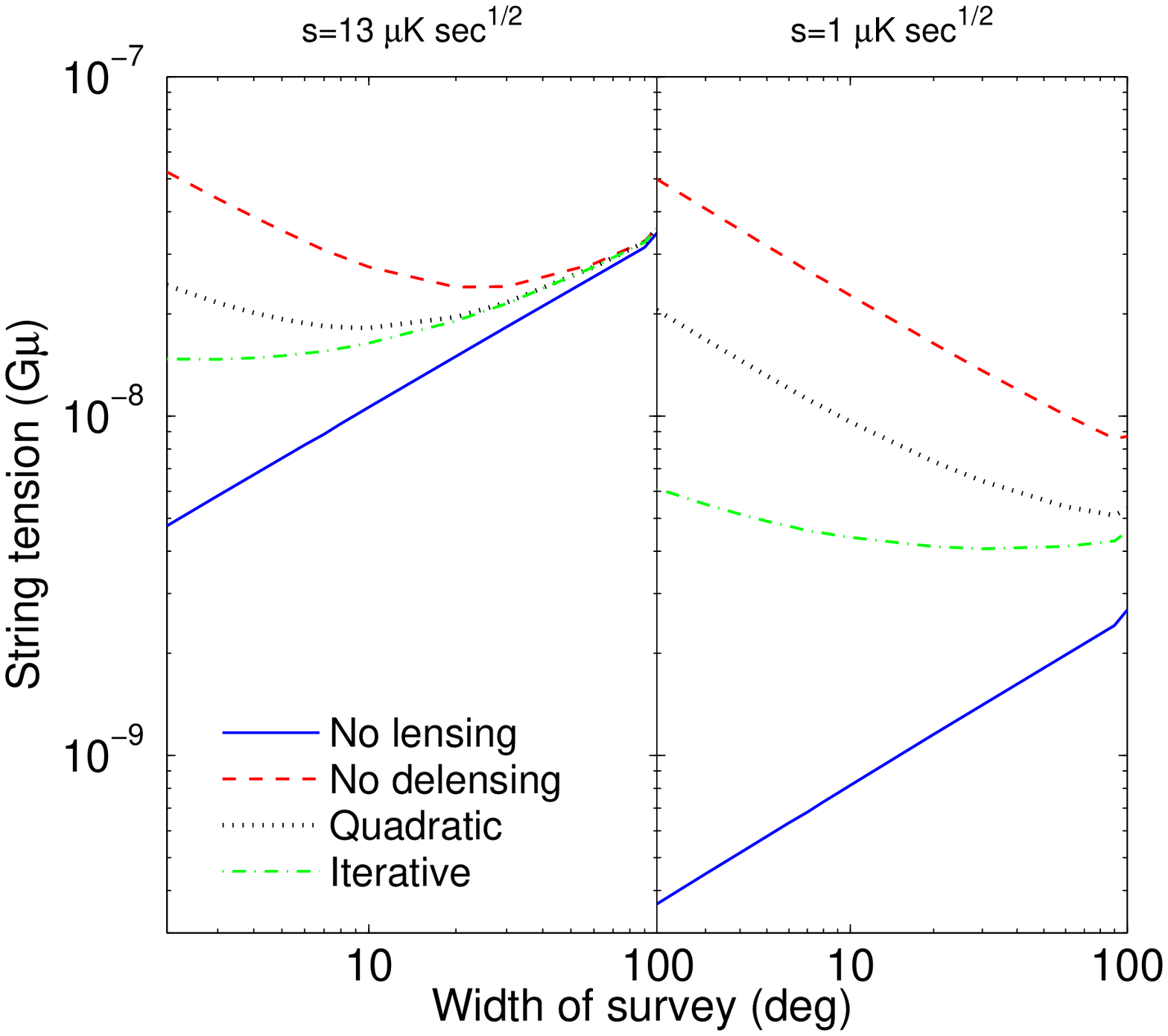}
\caption{The simple forecast for detecting tensor $B$-mode and cosmic strings. The horizontal axis is the sky coverage $\theta \sim  203f_{\rm sky}^{1/2}$. The vertical axis is smallest detectable tensor-to-scalar ratio $r$ (top) and strings tension $G\mu$ (bottom).  We set $s=13\mu$K\,sec$^{1/2}$ and $s=1\mu$K\,sec$^{1/2}$, 4 arcminute FWHM, and observation time $t_{\rm obs}=1$ year.}
\label{fig:plot4}
\end{figure}

For a given experimental survey speed it is important to optimize the survey width to obtain the best possible sensitivity on the  target signal.  We provide these ``experimentalists' guides" for both the tensor induced and cosmic string induced $B$-mode in Fig. \ref{fig:plot4}.  Plots similar to these exist in various forms for tensor-induced $B$-polarization, quantified by tensor-to-scalar  ratio $r$ \cite{kes02} \cite{jaffe00} \cite{coor03} \cite{lewis02}.  But to our knowledge, iterative estimators have never been considered in these plots, so we produce them here for comparison.  Astronomical foregrounds are also neglected as they can be removed with multi-frequency observations.  However, effects of cut sky have been ignored  so these plots still should not be regarded as final.  

In previous section, we showed that it is possible to delens at high $l$ region. The delensing algorithm formally only applies to Gaussian primordial signals. The rotation-generated $B$-mode is Gaussian.  Strictly speaking, both the string generated $B$-mode and the patchy reionization $B$-mode are non-Gaussian. As has been pointed out by \cite{dvor09}, it is possible to simultaneously reconstruct the $\phi$ field and the optical depth $\tau$ field. That would be compatible with the moving notch filtering approach proposed in the previous section. There is not enough information on the non-Gaussianity of string induced $B$-mode.  For demonstrative purpose only, we apply that method to  the search of string-induced $B$-mode by assuming Gaussianity.  Although cosmic strings do not source the  majority of initial perturbations, they are still generic predictions of particle physics and cosmology so should be searched for in  the data.  Some theoretical models such as KKLMMT \cite{kach03} can induce inflation along with cosmic strings.  Even the contributions from strings are severely limited by CMB temperature power spectrum, it has been shown in simulation that string  generated $B$-mode as large as the lensing induced $B$-mode is still allowed by data \cite{bevi07} (see Fig.\ref{fig:plot0}). 

For the stringy $B$-mode we assume the cosmic strings are smooth ($\alpha=1$) \cite{pogo08}, with segment RMS velocity $v_r=0.65$ and  initial correlation length divided by initial conformal time as $0.13$. 

When we fix the parameters other than $G\mu$, the power spectrum of $B$-mode will be proportional to $(G\mu) ^2$ so that we choose $(G\mu)^2$ to be our parameter for forecast. We use Fisher information matrix to be our estimate of ability to  constrain $(G\mu)^2$:

\begin{equation}
\label{eqn:fisher}
\sigma^{-2}_{(G\mu)^2}=\sum_{l>\frac{180}{\theta}}(\frac{\partial C_l^{BB,string}}{\partial (G\mu)^2})^2(\sigma^{BB}_l)^{-2},
\end{equation}  
where $\sigma_{(G\mu)^2}$ is Fisher limit for detecting quantity $(G\mu)^2$ (also the smallest detectable $(G\mu)^2$), $\sigma^{BB}_l$  is the standard error with which each multipole moment $C^{BB}_l$ can be determined.  Summing over $l>\theta/180$ makes sure no  information is coming from modes with wavelengths larger than the survey size (here we neglect $E$-$B$ mixing which will decrease the  number of useful modes).  We assume the delensed field will share the same covariance properties with a Gaussian field.  Since we  intend to do null test no string $B$-mode is included in $\sigma^{BB}_l$. Therefore,
\begin{equation}
\label{eqn:sigma} 
\sigma^{BB}_l=\sqrt{\frac{2}{f_{\rm sky}(2l+1)}}(C^{B_{\rm res}}_l+f_{\rm sky}\omega^{-1}e^{l(l+1)\sigma^2/8ln2}),
\end{equation}
where $f_{\rm sky}$ is the fraction of sky we survey. $\omega^{-1}=4\pi(s/T_{CMB})^2/t_{\rm obs}$ \cite{knox95}. Quantity $s$ is the  detector sensitivity in units $\mu$K\,sec$^{1/2}$, $\sigma$ is the FWHM of the beam and $t_{\rm obs}$ is the total observation time.   Here we set $s=13 \mu $K\,sec$^{1/2}$ and $t_{\rm obs}=1$ year (so that when $f_{\rm sky}\sim 0.0024$ or $\theta=10^{\circ}$,  $\Delta_p\sim 1.41\mu$K\,arcmin).

The forecast results are presented in Fig.\ref{fig:plot4}.

{\it Tensor Induced B-mode}   $\;\;\;$  When lensing is ignored, the $r$ vs. width curve recovers early predictions by \cite{jaffe00}, in which a survey width as small as $\sim 5^{\circ}$seems to be favored.  When lensing is treated as noise and no delensing is performed, the optimal survey width increases.  The minimum near or above 100 degrees is caused by the tensor reionization peak near $l\sim 6$ (Fig. \ref{fig:plot0}).  This apparent minimum drives some suborbital experiments to cover as much sky as possible at the risk of seeing the Galactic foreground emission.  With quadratic, and especially with iterative delensing, deep surveys become more  and more favorable as the instrumental noise decreases, although it never quite recovers the ``no lensing" curve.  Non-Gaussianity at  low noise level and $E$-$B$ leakage from cut-sky can further prevent the $r$ limit from reaching the predictions for iterative  delensing.  In addition to these factors, the choice of the survey width will be a practical balance between the required knowledge on noise properties and foregrounds (especially for wide surveys) and polarization/beam systematics (especially for deep surveys).  

{\it String induced B-mode} $\;\;\;$  The peak at $l\sim 600$ in the stringy $B$-mode spectrum (Fig. \ref{fig:plot0}) drives the optimal sky coverage  to $\theta < 20^{\circ}$ even without any delensing.  With delensing, the optimal coverage reduces further to $\theta \sim 10^{\circ}$.   In this case, the lens cleaning can improve the detectable $G\mu\sim 2.8\times 10^{-8}$ to $G\mu \sim 1.6\times 10^{-8}$ with $\theta \sim 10^{\circ}$ ($s=13\mu$K\,sec$^{1/2}$), $G\mu\sim 8\times 10^{-9}$ to $G\mu \sim 4\times 10^{-9}$ with $\theta \sim 100^{\circ}$ ($s=1\mu$K\,sec$^{1/2}$) .

\section{conclusion}

In contrary to previous assumptions in the literature, we show in this paper that many high order correlation terms introduced during  even the simplest delensing process are far from negligible.  This can be avoided by spatial filtering the observed $B$-map prior to  deflection field reconstruction, a practice that has been adopted for tensor- induced $B$-mode in \cite{selj04}.  We proposed a more  comprehensive fix to the problem.  With the new filtering scheme, delensing can be performed at high-$l$ region ($300<l<2000$) in a  simple manner.  We used simulations to verify the validity of the new process.  We assess how delensing improves $B$-mode searches  under different experimental parameters.  

\acknowledgments

WHT acknowledges the support of a visiting scholarship from the National Science Council, Taiwan.  CLK acknowledges the support of an  Alfred P. Sloan Research Fellowship.  

During preparation of the draft, we learned that an independent paper on the similar subject is being written by Dvorkin, Hanson, and  Smith, in which a different, more analytic fix will be presented.  We acknowledge useful  discussions with Kendrick Smith, especially his suggestion that this technique can be potentially useful for the study of patchy  reionization.  

\bibliographystyle{unsrt}

\end{document}